\documentclass[aps,preprint,floatfix,amsmath,superscriptaddress,endfloats,byrevtex]{revtex4}
\usepackage{graphicx}
\usepackage{feynmf}

\begin{document}

\title{Diagrammatic calculation of energy spectrum of quantum impurity in degenerate Bose-Einstein condensate}

\author{Alexey Novikov}
\author{Mikhail Ovchinnikov}
\affiliation{Department of Chemistry, University of Rochester, RC
Box 270216,
   Rochester, NY 14627-0216, USA}

\date{\today}

\begin{abstract}
In this paper we considered a quantum particle moving through delute
Bose-Einstein condensate at zero temperature. In our formulation the
impurity particle interacts with the gas of uncoupled Bogoliubov's
excitations. We constructed the perturbation theory for the Green's
function of the impurity particle with respect to the
impurity-condensate interaction employing the coherent-state path
integral approach. The perturbative expansion for the Green's
function is resumed into the expansion for its poles with the help
of the diagrammatic technique developed in this work. The dispersion
relation for the impurity clothed by condensate excitations is
obtained and effective mass is evaluated beyond the Golden rule
approximation.

\end{abstract}
\maketitle
\section{Introduction}
\label{intro} Significant number of theoretical works were devoted
to the quantum theory of superfluidity on a microscopic scale.  The
recent interest stems from several new experiments on the superfluid
Helium and other Bose-Einstein condensates (BEC). Rotational motion
of molecules has been extensively studied in the superfluid helium
droplets \cite{vilesov1998, vilesov2000}. A unique properties of the
measured rotational spectra pose a large number questions, such as:
what is the collective molecule/superfluid wavefunction that
describes sharp rotational states observed in experiments; what are
the properties of finite systems and how is the limit of a bulk
superfluid is reached \cite{dumesh2006}?  A similar microscopic
phenomena were studied in the BEC of Sodium atoms in magnetic traps
\cite{chikkatur2000}. A linear motion of impurities was shown to be
dissipationless for the speeds below the condensate speed of sound.
A large number of theoretical works addressed the molecule-He
droplet system using imaginary time path integral Monte-Carlo
approaches \cite{dalfovo2001, whaley2004p, whaley2001k, whaley2004z,
whaley2004zk, whaley2003p, roy2006}. While in certain cases
remarkable agreement with experimental constants was obtained
\cite{roy2006}, those works are strictly limited to the calculation
of statistical properties and thus provides no real understanding of
the microscopic nature of the dissipationless motion.  The latter
can only be established by considering a real time dynamics. A
number of theoretical works considered a motion of impurity through
delute BEC.  A macroscopic particle interacting with delute BEC has
been considered \cite{suzuki2005}. In this case the motion of
particle is equivalent to the BEC in a time dependent external
potential. This problem was treated by solving time-dependent
Gross-Pitaevskii equations \cite{astrak2004}. A microscopic particle
interacting with the delute BEC in Bogoliubov's approximation has
been considered by several authors using general Golden rule
considerations \cite{montina2003, montina2002, timmermans98}. These
works were based on the result of Miller et al \cite{miller62} which
was obtained using time-independent perturbation theory. The
Bogoliubov's treatment has also been successfully used for the
investigation of the force acting on the impurity particle due to
the quantum fluctuations in BEC \cite{roberts2006, roberts2005}.
Some authors treated a particle strongly interacting with
Bogoliubov's BEC and found a possibility of self localization
\cite{kalas2006, cucchietti2006, sacha2006}. In summary, all of the
previous works dealt with the lowest order of the perturbation
theory, either using Golden rule approximation or considering the
interaction with the fluctuations of the Gross-Pitaevskii field.
What we seek is the perturbation theory that can be systematically
extended to an arbitrary order.  In present paper we use the field
theoretical methods to develop such treatment.

In our recent work \cite{novikov2008gr} we described an impurity
moving through BEC as a microscopic particle within time-dependent
perturbation theory. We introduced the Hamiltonian of a quantum
particle moving within the interacting Bose gas. No assumption was
made about the relative mass of an impurity compared to that of the
Bose particles. After the introduction of the general Hamiltonian,
the Bogoliubov's approximation was made to convert the Hamiltonian
to the diagonal form. Then the problem was reduced to the quantum
particle moving through the gas of non-interacting Bogoliubov's
excitations.  The natural way to compute the real time dynamics of
impurity is to develop an expansion considering particle-BEC
interaction as a perturbation.  Our previous as well several works
of other authors delt with the lowest order term that corresponds to
the Golden rule limit. The higher order terms in direct perturbative
expansion are diverging with time. Thus direct perturbation theory
does not work beyond the Golden rule limit.

In this work we propose a method which allows to avoid such
divergence and consider higher orders of perturbation theory. The
time evolution is written in terms of coherent state path integral.
Because of the linear dependence of the interaction Hamiltonian on
the BEC degrees of freedom this functional integral can be reduced
to the non-Gaussian integral over the particle trajectories. The
non-Gaussian part describing particle-BEC interaction leads to the
formal perturbation expansion. In order to prevent the appearance of
the secular terms, the perturbation series is resummed as an
expansion for the poles of the Green's function with the help of the
diagrammatic technique developed in this work. In this way we obtain
the expression for the self-energy operator which is responsible for
the shift of the Green's function pole due to the particle-BEC
interaction, i.e. the dispersion relation for the impurity dressed
by the cloud of BEC excitations. The limit of the self energy at
small momenta allows to compute the effective mass of the particle
up to an arbitrary order of the perturbation theory. As an example
of this technique we compute the term of the perturbation expansion
that is next to the Golden rule result and its contribution to the
energy spectrum and effective mass of the particle.

\section{MODEL HAMILTONIAN}

\label{model}

Let us start with the Hamiltonian of interacting Bose particles in
secondary quantization representation
\begin{equation}
\label{ham2} H_B=\sum_{\bf p}\frac{{\bf p}^2}{2m}b^+_{{\bf
p}}b_{{\bf p}}+\frac{1}{2V}\sum_{{\bf p}_1,{\bf p}_2,{\bf p}}U({\bf
p})b^+_{{\bf p}_1-{\bf p}}b^+_{{\bf p}_2+{\bf p}}b_{{\bf
p}_2}b_{{\bf p}_1}~,
\end{equation}
where $U({\bf p})$ is the Fourier transform of the interaction
potential
\begin{equation}
\label{potential} U({\bf p})=\int U({\bf r})e^{i{\bf pr}}d{\bf r}~.
\end{equation}
The Plank constant $\hbar$ is set to unity here and throughout the
paper. We will concentrate on the case of delute gas $r_0\ll
n^{-1/3}$ where $r_0$ is the range of potential on which $U(r)$
differs from zero significantly and $n$ denotes density of gas. So
the Fourier transform of the interaction potential $U({\bf p})$ can
be replaced by its zero component $U_0=\int\d{\bf r}U({\bf r})$
which is connected with the length of $s$-scattering in first order
Born approximation as following
\begin{equation}
a_s=\frac{m}{4\pi}U_0~.
\end{equation}
Then we will consider degenerate gas at zero temperature. In this
case the Hamiltonian ({\ref{ham2}) can be reduced to the diagonal
form with the help of the Bogoliubov's method \cite{bogol1947}.
\begin{equation}
\label{ham4} H_B=E_0+\sum_{\bf p}\epsilon(p)B^+_{\bf p}B_{\bf p}~.
\end{equation}
Here the new bosonic operators $B_{\bf p}^+$ and $B_{\bf p}$ create
and annihilate the collective excitations in BEC with the spectrum
\begin{equation}
\epsilon(p)=\sqrt{\frac{p^2}{2m}\left(\frac{p^2}{2m}+2nU_0\right)}~,
\end{equation}
which has the phonon-like behavior at low momenta, i.e.
$\epsilon(p\to0)=p\sqrt{nU_0/m}=pc$, where $c$ is the speed of
sound. The original particle operators $b_{\bf p}^+$ and $b_{\bf p}$
are connected with the operators of Bogoliubov's excitations $B_{\bf
p}^+$ and $B_{\bf p}$ by following relations \cite{landau_v9}
\begin{equation}
\label{transform} b_{\bf p}=\alpha_{\bf p}B_{\bf p}+\beta_{\bf
p}B^+_{\bf -p}~,~~~
\\
b^+_{\bf p}=\alpha_{\bf p}B^+_{\bf p}+\beta_{\bf p}B_{\bf -p}~.
\end{equation}
where the transformation coefficients are
\begin{eqnarray}
\alpha_{\bf p}&=&\frac{\mu_{\bf p}}{\sqrt{\mu_{\bf
p}^2-1}}~,~~~\beta_{\bf p}=\frac{1}{\sqrt{\mu_{\bf p}^2-1}}~,~~~
\\
\mu_{\bf p}&=&-\frac{\epsilon({\bf
p})+p^2/2m+nU_0}{nU_0}~\nonumber~.
\end{eqnarray}
The ground state energy of BEC is given by
\begin{equation}
\label{gsebec}
E_0=\frac{nU_0}{2}(N-1)+\frac{1}{2}\sum_{{\bf p}\not=
0}\left(\epsilon(p)-\frac{p^2}{2m}-nU_0\right)~.
\end{equation}

Next, we will consider a single quantum particle with mass $M$ and
momentum ${\bf q}$ interacting with the environment of Bose gas
discussed above. The whole system is then described by following
Hamiltonian
\begin{equation}
\label{ham5} H=\sum_{\bf q}\frac{q^2}{2M}a_{\bf q}^+a_{\bf
q}+H_B+H_{I}
\end{equation}
with the particle-environment interaction
\begin{equation}
\label{ham6} H_{I}=\frac{g}{V}\sum_{{\bf p,k,q}}b_{{\bf p}+{\bf
k}}^+b_{\bf p}a^+_{{\bf q}-{\bf k}}a_{\bf q}~.
\end{equation}
Here the bosonic operators $a^+_{\bf q}/a_{\bf q}$ create/annihilate
the particle in state $|{\bf q}\rangle$ and the coupling constant
$g$ is determined as zero Fourier component of the
system-environment interaction. After application of the
Bogoluibov's transformation to the interaction (\ref{ham6}) the
Hamiltonian of the whole system (\ref{ham5}) takes the form
\begin{eqnarray}
\label{fullham} H&=&E^\prime_0 +\sum_{\bf q}\frac{{\bf
q}^2}{2M}a_{\bf q}^+a_{\bf q}+\sum_{\bf p}\epsilon(p)B^+_{\bf
p}B_{\bf p}+\sum_{{\bf q},{\bf p}\not= 0}\gamma_{\bf
p}\left(a^+_{{\bf q}-{\bf p}}a_{\bf q}B^+_{\bf p}+a^+_{{\bf q}+{\bf
p}}a_{\bf q}B_{\bf p}\right)~,
\\
\gamma_{\bf
p}&=&\frac{g}{V}\sqrt{\frac{Np^2}{2m\epsilon(p)}}~.\nonumber
\end{eqnarray}
The ground state energy $E_0+gn$ is now shifted with respect to the
$E_0$ due to the system-condensate interaction.

\section{COHERENT-STATE PATH INTEGRAL FORMULATION OF EVOLUTION OPERATOR}

\label{cspi}

The main task of present work is the evaluation of the dynamical
quantities such as transition amplitudes or Greens' function of the
system described the Hamiltonian in bosonic creation/annihilation
operator representation. In order to proceed with such kind of
calculations one can employ the coherent state path integral
technique \cite{novikov2004, baranger2001}. In this section we will
give some basic introduction into the coherent state path integral
formulation of the dynamics of the Bose many-particle systems.

First, let us define the coherent state of many-particle Bose system
\begin{eqnarray}
|\{z_{\bf p}\}\rangle=|z_{{\bf p}_1},z_{{\bf p}_2},...,z_{{\bf
p}_k},...\rangle
\end{eqnarray}
as an eigenstate of annihilation operator
\begin{eqnarray}
\hat{a}_{{\bf p}_k}|\{z_{\bf p}\}\rangle=z_{{\bf p}_k}|\{z_{\bf
p}\}\rangle~.
\end{eqnarray}
One can write the matrix element of the evolution operator in
coherent state basis as the following functional integral
\begin{eqnarray}
\label{evolution}
\label{funcint}
\langle \{z_{\bf
p}\}|e^{-iHt}|\{z_{\bf p}^{\prime}\}\rangle&=&\int D[\{z^{\ast}_{\bf
p}(\tau)\},\{z_{\bf p}(\tau)\}]
\\
&\times&\exp\left[-\frac{1}{2}\sum_{\bf p}(|z_{\bf
p}|^2+|z^{\prime}_{\bf p}|^2) -i S(\{z_{\bf
p}(\tau)\},\{z^{\ast}_{\bf p}(\tau)\})\right]~.\nonumber
\end{eqnarray}
Here $S$ denotes the action depending on the trajectories $z_{\bf
p}(\tau)$ and  $ z^{\ast}_{\bf p}(\tau)$
\begin{eqnarray}
\label{act1} S(\{z_{\bf p}(\tau)\},\{z^{\ast}_{\bf p}(\tau)\})=
i\sum_{\bf p}z^{\ast}_{\bf p}(t)z_{\bf p}(t)+i\sum_{\bf p}z_{\bf
p}^\ast(\tau)\dot{z}_{\bf p}(\tau)+H(\{z_{\bf
p}^\ast(\tau)\},\{z_{\bf p}(\tau)\})~,
\end{eqnarray}
which must be evaluated with the boundary conditions
\begin{eqnarray}
z_{\bf p}(0)=z_{\bf p}^{\prime}  ~, ~~ z^{\ast}_{\bf
p}(t)=z^{\ast}_{\bf p} ~,
\end{eqnarray}
where $z^{\ast}_{\bf p}$ and $z_{\bf p}^{\prime}$ correspond to the
bra and ket states on the left hand side of Eq.~(\ref{evolution}),
respectively. The integration in (\ref{evolution}) is performed over
all trajectories depending on time, and the symbol $D[...]$ is
determined as
\begin{eqnarray}
D[\{z^{\ast}_{\bf p}(\tau)\},\{z_{\bf p}(\tau)\}]=\prod_{\bf
p}\prod_\tau\frac{dz^\ast_{\bf p}(\tau)dz_{\bf p}(\tau)}{\pi}~.
\end{eqnarray}
Here we have to note that the trajectories $z^{\ast}_{\bf p}(\tau)$
and $z_{\bf p}(\tau)$ are different functions and are not conjugated
of each other.

Now let us concentrate on the vacuum amplitude for free Bose gas
described by the Hamiltonian $H=\sum_{\bf p}E(p)a^+_{\bf p}a_{\bf
p}$
\begin{eqnarray}
\label{vacam}
\langle 0|\exp\big(-it\sum_{\bf p}E(p)a^{+}_{\bf
p}a_{\bf p}\big)|0\rangle=\int D[\{z^{\ast}_{\bf p}(\tau)\},\{z_{\bf
p}(\tau)\}]e^{-iS_0}~,
\end{eqnarray}
where $S_0$ is the free action
\begin{eqnarray}
S_0=\sum_{\bf p}\int_0^td\tau z^{\ast}_{\bf p}(\tau)\hat{G}_{\bf
p}z_{\bf p}(\tau)~,~~~ \hat{G}_{\bf
p}=-i\frac{\partial}{\partial\tau}+E(p)~.
\end{eqnarray}
Since the vacuum state is defined as $|0\rangle=|\{z_{\bf
p}\}=0\rangle$, the functional integral (\ref{vacam}) has to be
evaluated with zero boundary conditions, i.e. $z_{\bf
p}(0)=z^\ast_{\bf p}(t)=0$. Below we will need the so called
generating functional for the vacuum amplitude which is defined by
introducing some auxiliary sources $j^\ast_{\bf p}(\tau)$ and
$j^\ast_{\bf p}(\tau)$ into the action, i.e.
\begin{eqnarray}
\label{gfunctional0} \Lambda[\{j_{\bf p}^\ast(\tau)\},\{j_{\bf
p}(\tau)\}]&=&\int D[\{z^{\ast}_{\bf p}(\tau)\},\{z_{\bf p}(\tau)\}]
\nonumber
\\
&\times&\exp\left[-iS_0+\int_0^td\tau\sum_{\bf p}\big(z^\ast_{\bf
p}(\tau)j^\ast_{\bf p}(\tau)+z_{\bf p}(\tau)j_{\bf
p}(\tau)\big)\right]~.
\end{eqnarray}
Note that the sources $j_{\bf p}^\ast(\tau)$ and $j_{\bf p}(\tau)$
as well as the trajectories $z^{\ast}_{\bf p}(\tau)$ and $z_{\bf
p}(\tau)$ are different functions, so they are not conjugated. The
above functional integral (\ref{gfunctional0})can be evaluated by
introducing the new integration variables
\begin{eqnarray}
\label{deviations}
z_{\bf p}(\tau)&=&\bar{z}_{\bf p}(\tau)+\delta
z_{\bf p}(\tau)~,\nonumber
\\
z^\ast_{\bf p}(\tau)&=&\bar{z}^\ast_{\bf p}(\tau)+\delta z^\ast_{\bf
p}(\tau)~,
\end{eqnarray}
where the stationary trajectories $\bar{z}^\ast_{\bf p}(\tau)$ and
$\bar{z}_{\bf p}(\tau)$ are determined by following equations
\begin{eqnarray}
\label{stattreq}
\hat{G}_{\bf p}\bar{z}_{\bf p}(\tau)+j_{\bf
p}^\ast(\tau)&=&0~,\nonumber
\\
\hat{G}^+_{\bf p}\bar{z}^\ast_{\bf p}(\tau)+j_{\bf p}(\tau)&=&0~.
\end{eqnarray}
Substituting trajectories in the form (\ref{deviations}) into the
integral (\ref{gfunctional0}) and using Eqs.~(\ref{stattreq}), for
the generation functional one gets
\begin{eqnarray}
\label{gfunctional1}
\Lambda[\{j_{\bf p}^\ast(\tau)\},\{j_{\bf
p}(\tau)\}]&=&\exp\left(\int_0^td\tau\sum_{\bf p}z_{\bf
p}(\tau)j_{\bf p}(\tau)\right)
\\
&\times&\int D[\{\delta z^{\ast}_{\bf p}(\tau)\},\{\delta z_{\bf
p}(\tau)\}]\exp\left(\int_0^td\tau\sum_{\bf p}\delta z^\ast_{\bf
p}(\tau)\hat{G}_{\bf p}\delta z_{\bf p}(\tau)\right)~.\nonumber
\end{eqnarray}
The remaining integral over deviations $\delta z^{\ast}_{\bf
p}(\tau)$ and $\delta z_{\bf p}(\tau)$ in Eq.~(\ref{gfunctional1})
exactly corresponds to the vacuum amplitude for free Bose particles
described by normally ordered Hamiltonian and hence equals unity.
Finally, after substituting the solution of the equation for the
stationary trajectories (\ref{stattreq}), generating functional
takes the form
\begin{eqnarray}
\label{gfunctional2} \Lambda[\{j_{\bf p}^\ast(\tau)\},\{j_{\bf
p}(\tau)\}]&=&\exp\left(\sum_{\bf
p}\int_0^td\tau\int_0^td\tau^\prime G_{\bf
p}(\tau-\tau^\prime)j_{\bf p}(\tau)j_{\bf
p}^\ast(\tau^\prime)\right)~,
\end{eqnarray}
where $G_{\bf p}(\tau)$ is the Green's function of the operator
$\hat{G}_{\bf p}$
\begin{eqnarray}
\label{green0}
 G_{\bf
p}(\tau-\tau^\prime)&=&\Theta(\tau-\tau^\prime)\exp[-iE(p)(\tau-\tau^\prime)]~,
\end{eqnarray}
and $\Theta(\tau)$ denotes the Heaviside step function.

The last point is to determine the mean value of some functional of
trajectories ${\mathcal A}[\{z_{\bf p}(\tau)\},\{z^\ast_{\bf
p}(\tau)\}]$ as follows
\begin{eqnarray}
\langle {\mathcal A}\rangle=\int D[\{z^{\ast}_{\bf
p}(\tau)\},\{z_{\bf p}(\tau)\}]{\mathcal A}e^{-iS_0}~.
\end{eqnarray}
In the next section we will have to evaluate the mean products of
the trajectories taken in different moments of time which can be
written as the functional derivative of the generating functional
\begin{eqnarray}
\label{funcderiv}
 &&\langle z_{{\bf p}_1}(s_1)z_{{\bf
p}_2}(s_2)...z_{{\bf p}_n}(s_n)z^\ast_{{\bf
p}_{n+1}}(s_{n+1})z^\ast_{{\bf p}_{n+2}}(s_{n+2})...z^\ast_{{\bf
p}_{2n}}(s_{2n})\rangle= \nonumber
\\
&&\left.\frac{\delta^{2n}}{\delta j_{{\bf p}_1}(s_1)\delta j_{{\bf
p}_2}(s_2)...\delta j_{{\bf p}_n}(s_n)\delta j^\ast_{{\bf
p}_{n+1}}(s_{n+1})\delta j^\ast_{{\bf p}_{n+2}}(s_{n+2})...\delta
j^\ast_{{\bf p}_{2n}}(s_{2n})}\right|_{j_{\bf p}=j^\ast_{\bf p}=0}
\nonumber
\\
&&\times\Lambda[\{j_{\bf p}^\ast(\tau)\},\{j_{\bf p}(\tau)\}]~.
\end{eqnarray}
For example, the one particle Green's function reads
\begin{eqnarray}
\langle z_{{\bf p}}(s)z^\ast_{{\bf
p^\prime}}(s\prime)\rangle=\left.\frac{\delta^2\Lambda}{\delta
j_{\bf p}(s)\delta j^\ast_{\bf p^\prime}(s^\prime)}\right|_{j_{\bf
p}=j^\ast_{\bf p}=0}=G_{\bf p}(s-s^\prime)\delta_{{\bf p},{\bf
p^\prime}}~.
\end{eqnarray}
Since in the case $s=s^\prime$ the Green's function $G(0)=\langle
a^+(s)a(s)\rangle=0$, the expression (\ref{green0}) must be written
in the form
$G(s-s^\prime)=\Theta(s-s^\prime-0)\exp[-iE(p)(s-s^\prime)]$.

\section{DIAGRAMMATIC TECHNIQUE FOR THE GREEN'S FUNCTION OF THE RELEVANT PARTICLE}
The purpose of this section is the construction of the perturbation
theory for the Green's function of the impurity particle ${\mathcal
G}_{\bf p}(t)$ defined as the correlation function of the creation
and annihilation operators, i.e. ${\mathcal G}_{\bf p}(t)=\langle
a_{\bf p}(t)^+a_{\bf p}(0)\rangle$. Let us start with the transition
amplitude
\begin{eqnarray}
w_{i\to f}=\langle 0|_B\langle f|e^{-iHt}|i\rangle|0\rangle_B~,
\end{eqnarray}
which describes the transition of the impurity particle from some
initial state $|i\rangle$ to some final state $|f\rangle$ while the
BEC remains in its vacuum state $|0\rangle_B$, i.e. state with an
absence of Bogoliubov's excitations. It is clear that in this case
$|i\rangle=|f\rangle$. Below we will consider the eigenstate of
momentum $|{\bf p}\rangle$ as the initial and final states of the
impurity
\begin{eqnarray}
\label{tram}
 w_{i=f=|{\bf p}\rangle}={\mathcal G}_{\bf p}=\langle
0|_B\langle{\bf p}|e^{-iHt}|{\bf p}\rangle|0\rangle_B~.
\end{eqnarray}
One can write the transition amplitude as the following correlation
function
\begin{eqnarray}
{\mathcal G}_{\bf p}(t)=\langle 0|_B\langle 0|a_{\bf
p}e^{-iHt}a^+_{\bf p}|0\rangle|0\rangle_B=\langle a_{\bf p}(t)a_{\bf
p}^+(0)\rangle~.
\end{eqnarray}
Here $|0\rangle$ denotes the state of the impurity with no particle
and the brackets $\langle ...\rangle$ mean averaging over vacuum
states of the BEC and the impurity. Thus we see that the transition
amplitude of the form (\ref{tram}) coincides with the Green's
function of the impurity particle, and its poles determine the
excitation spectrum of the particle interacting with the surrounding
BEC. Now we can employ the path integral formulation of the matrix
element of evolution operator developed in Sec.~\ref{cspi}
\begin{eqnarray}
\label{func01} &&{\mathcal G}_{\bf p}(t)=\int D[\{a^{\ast}_{\bf
q}(\tau)\},\{a_{\bf q}(\tau)\}]~a_{\bf p}(t)a^\ast_{\bf p}(0)\int
D[\{b^{\ast}_{\bf q}(\tau)\},\{b_{\bf q}(\tau)\}]\nonumber
\\
&&\times\exp\left[-iS_P-iS_B-i\int_0^td\tau\sum_{{\bf
q,q^\prime}\not=0}\gamma_{\bf q}\big(a^\ast_{\bf
q^\prime-q}(\tau)a_{\bf q^\prime}(\tau)b^\ast_{\bf
q}(\tau)+a^\ast_{\bf q^\prime+q}(\tau)a_{\bf q^\prime}(\tau)b_{\bf
q}(\tau)\big)\right]~,
\end{eqnarray}
where we have determined the action of the free impurity particle
\begin{eqnarray}
\label{sp}
 S_P=\sum_{\bf q}\int_0^td\tau a^{\ast}_{\bf
q}(\tau)\hat{G}_{\bf q}a_{\bf q}(\tau)~, ~~~ \hat{G}_{\bf
q}=-i\frac{\partial}{\partial\tau}+\frac{q^2}{2M}~,
\end{eqnarray}
and the action of free BEC
\begin{eqnarray}
S_B=\sum_{\bf q}\int_0^td\tau b^{\ast}_{\bf
q}(\tau)\hat{\Gamma}_{\bf q}b_{\bf q}(\tau)~, ~~~ \hat{\Gamma}_{\bf
q}=-i\frac{\partial}{\partial\tau}+\epsilon(q)~.
\end{eqnarray}
Since the interaction part of the action in the functional integral
(\ref{func01}) has linear dependence on the $b^\ast(\tau), b(\tau)
$-trajectories, the BEC degrees of freedom can be immediately
integrated out in the same manner as with the calculation of the
generating functional (\ref{gfunctional2}).

After eliminating the BEC from (\ref{func01}), for the correlation
function ${\mathcal G}_{\bf p}$ one gets
\begin{eqnarray}
\label{nGfi}
{\mathcal G}_{\bf p}(t)=\int D[\{a^{\ast}_{\bf
q}(\tau)\},\{a_{\bf q}(\tau)\}]~a_{\bf p}(t)a^\ast_{\bf
p}(0)\exp\left[-iS_P-iS_I\right]~,
\end{eqnarray}
The above integral is the non-Gaussian functional integral over
impurity particle trajectories only, and its non-Gaussian part $S_I$
reads
\begin{eqnarray}
S_I=\int_0^td\tau\int_0^td\tau^\prime\sum_{q,q^\prime,q^{\prime\prime}}a^\ast_{\bf
q^\prime+q}(\tau)a_{\bf q^\prime}(\tau)\Gamma_{\bf
q}(\tau-\tau^\prime)a^\ast_{\bf
q^{\prime\prime}-q}(\tau^\prime)a_{\bf
q^{\prime\prime}}(\tau^\prime)~,
\end{eqnarray}
Here $\Gamma_{\bf q}(\tau)$ represents the Green's function of the
operator $\hat{\Gamma}_{\bf q}$. Now our aim is to construct the
perturbative expansion for the path integral of the form
(\ref{nGfi}) in powers of its non-Gaussian part
\begin{eqnarray}
{\mathcal G}^{(n)}_{\bf p}(t)=\langle a_{\bf p}(t)a^\ast_{\bf
p}(0)\rangle-\langle a_{\bf p}(t)S_Ia^\ast_{\bf
p}(0)\rangle+\frac{1}{2!}\langle a_{\bf p}(t)S_I^2a^\ast_{\bf
p}(0)\rangle+...+\frac{(-1)^n}{n!}\langle a_{\bf
p}(t)S_I^na^\ast_{\bf p}(0)\rangle~.
\end{eqnarray}
The general expression for the n-th term of the above expansion
reads
\begin{eqnarray}
\label{gexprG} &&\langle a_{\bf p}(t)S_I^na^\ast_{\bf
p}(0)\rangle=\sum_{\bf
q_1,q_2,...,q_n}\int_0^td\tau_1d\tau_2...d\tau_n\int_0^td\tau_1^\prime
d\tau_2^\prime...d\tau_n^\prime\Gamma_{\bf
q_1}(\tau_1-\tau_1^\prime)\Gamma_{\bf
q_2}(\tau_2-\tau_2^\prime)...\Gamma_{\bf
q_n}(\tau_n-\tau_n^\prime)\nonumber
\\
&&\times\sum_{\bf q^\prime_1,q^\prime_2,...,q^\prime_n}\sum_{\bf
q^{\prime\prime}_1,q^{\prime\prime}_2,...,q^{\prime\prime}_n}
\langle a_{\bf p}(t)a^\ast_{\bf q_1^\prime+q_1}(\tau_1)a_{\bf
q_1^\prime}(\tau_1)a^\ast_{\bf
q_1^{\prime\prime}-q_1}(\tau_1^\prime)a_{\bf
q_1^{\prime\prime}}(\tau_1^\prime)...\nonumber
\\
&&~~~~~~~~~~~~~~~~~~~~~~~~~~\times a^\ast_{\bf
q_n^\prime+q_n}(\tau_n)a_{\bf q_n^\prime}(\tau_n)a^\ast_{\bf
q_n^{\prime\prime}-q_n}(\tau_n^\prime)a_{\bf
q_n^{\prime\prime}}(\tau_n^\prime)a_{\bf p}^\ast(0)\rangle ~,
\end{eqnarray}
where the averaging is performed by the integration over all
trajectories with the weight $\exp(-iS_P)$.

First, let us consider the correlation functions of the trajectories
$a^\ast_{\bf q}(\tau)$ and $a_{\bf q}(\tau)$ in the integrand of
Eq.~(\ref{gexprG}). Using general formula (\ref{funcderiv}) together
with the expression for the generating functional for the impurity
particle in the form of Eq.~(\ref{gfunctional2}) where the
propagator $G_{\bf q}(\tau)$ is determined as the Green's function
of the operator ${\hat G}_{\bf q}$ in Eq.~(\ref{sp}), for the
correlator of the particle trajectories one gets
\begin{eqnarray}
\label{corr1}
&&\langle
a(t)a^\ast(\tau_1)a(\tau_1)a^\ast(\tau_1^\prime)a(\tau_1^\prime)a^\ast(\tau_2)a(\tau_2)a^\ast(\tau_2^\prime)a(\tau_2^\prime)
...a^\ast(\tau_n)a(\tau_n)a^\ast(\tau_n^\prime)a(\tau_n^\prime)a^\ast(0)\rangle\nonumber
\\
&&=\sum_{P(\{\tau_k\},\{\tau_k^\prime\})}G(t-P_1)G(P_1-P_2)...G(P_{2n-1}-P_{2n})G(P_{2n})~.
\end{eqnarray}
We omitted the momentum indexes in the above expression for
simplicity. The sum in the right hand side of Eq.~(\ref{corr1}) is
performed over all permutation of the time points
$\tau_1,\tau_2,...,\tau_n,\tau^\prime_1,\tau^\prime_2,...,\tau^\prime_n$.
This equation which we obtained using the method of generating
functional is in fact equivalent to the well known Wick's theorem.
Next, we can substitute the equation (\ref{corr1}) into the
expression (\ref{gexprG}) and do the following: instead of permuting
the time points in the correlator (\ref{corr1}) we will permute the
time points in the product of the BEC propagators
$\Gamma(\tau_1-\tau_1^\prime)\Gamma(\tau_2-\tau_2^\prime)...\Gamma(\tau_n-\tau_n^\prime)$
in the integrand in the right hand side of Eq.~(\ref{gexprG}) while
the particle correlator has to be taken with the fixed times
$\tau_k, \tau_k^\prime$, i.e.
\begin{eqnarray}
\label{gexprG2} &&\langle a_{\bf p}(t)S_I^na^\ast_{\bf
p}(0)\rangle=\int_0^td\tau_1d\tau_2...d\tau_n\int_0^td\tau_1^\prime
d\tau_2^\prime...d\tau_n^\prime
\sum_{P(\{\tau_k\},\{\tau_k^\prime\})}\Gamma(P_1-P_2)\Gamma(P_2-P_4)...\Gamma(P_{2n-1}-P_{2n})\nonumber
\\
&&\times
G(t-\tau_1)G(\tau_1-\tau_1^\prime)G(\tau_1^\prime-\tau_2)G(\tau_2-\tau_2^\prime)...G(\tau_{n-1}^\prime-\tau_n)G(\tau_n-\tau_n^\prime)G(\tau_n^\prime)~.
\end{eqnarray}

Now we are able to represent each term of perturbative expansion of
the Green's function ${\mathcal G}(t)$ graphically with the help of
Feynman's diagrams. Let us represent the product of the particle
propagators by the solid lines connecting the time points

\begin{eqnarray}
\label{gggdiag}
G(t-\tau_1)G(\tau_1-\tau_1^\prime)G(\tau_1^\prime-\tau_2)...G(\tau_n-\tau_n^\prime)G(\tau_n^\prime)=
\parbox{37mm}{
\begin{fmffile}{diag0001}
\begin{fmfgraph}(100,-0)
\fmfleft{o} \fmf{plain}{i,v1} \fmf{plain}{v1,v2} \fmf{plain}{v2,v3}
\fmf{plain}{v3,o} \fmfdot{v1,v2,v3} \fmfright{i}
\end{fmfgraph}
\end{fmffile}
}\cdot\cdot\cdot
\parbox{27mm}{
\begin{fmffile}{diag0003}
\begin{fmfgraph}(70,-0)
\fmfleft{o} \fmf{plain}{i,v1} \fmf{plain}{v1,v2} \fmf{plain}{v2,o}
\fmfdot{v1,v2} \fmfright{i}
\end{fmfgraph}
\end{fmffile}
}~
\end{eqnarray}
The number of vertices equals $2n$ where $n$ is the order of
perturbation. The right incoming and left outgoing plain lines
correspond to the trajectories $a_{\bf p}^\ast(0)$ and $a_{\bf
p}(t)$ in the expression for the correlator (\ref{corr1}),
respectively. Thus the zeroth order term of expansion is simply
given by
\begin{eqnarray}
{\mathcal G}^{(0)}_{\bf p}(t)=\langle a_{\bf p}(t)a_{\bf
p}^\ast(0)\rangle=G_{\bf p}(t)=
\parbox{37mm}{
\begin{fmffile}{diag0005}
\begin{fmfgraph}(50,-0)
\fmfleft{o} \fmf{plain}{i,o}  \fmfright{i}
\end{fmfgraph}
\end{fmffile}
}~
\end{eqnarray}
Then each pair of vertices has to be connected by the BEC propagator
$\Gamma$ by the all possible ways is accordance with the permutation
of the time points. We will represent the propagators $\Gamma$ by
wiggly lines. Let us illustrate this technique with the example of
the first order expansion term
\begin{eqnarray}
{\mathcal G}^{(1)}_{\bf p}(t)&=&{\mathcal G}^{(0)}_{\bf
p}(t)-\int_0^td\tau\int_0^td\tau^\prime\sum_{\bf q}G_{\bf
p}(t-\tau)G_{\bf p-q}(\tau-\tau^\prime)\Gamma_{\bf
q}(\tau-\tau^\prime)G_{\bf p}(\tau) \nonumber
\\
\nonumber \\
 &=&
\parbox{23mm}{
\begin{fmffile}{diag7250112}
\begin{fmfgraph}(60,-0)
\fmfleft{o} \fmf{plain}{i,o} \fmfright{i}
\end{fmfgraph}
\end{fmffile}}
-~
\parbox{37mm}{
\begin{fmffile}{diag0461}
\begin{fmfgraph}(90,-0)
\fmfleft{o} \fmf{plain}{i,v1} \fmf{phantom, right}{i,v1}
\fmf{plain}{v1,v2} \fmf{plain}{v2,o} \fmf{phantom, right}{v2,o}
\fmf{wiggly,right}{v1,v2} \fmfdot{v1,v2} \fmfright{i}
\end{fmfgraph}
\end{fmffile}
}
\end{eqnarray}

Each propagator line implies the sum over its momentum, while each
vertex corresponds to the time point and integration over it.
Besides, the vertex insures the momentum conservation rule, i.e. the
sum of momenta of all incoming lines equals the sum of momenta of
all outgoing lines.

Due to the property of the Green's function $G(s)\sim\Theta(s)$ in
the product on the right hand side of Eq.~(\ref{gggdiag}) we have to
set
$t>\tau_1>\tau_1^\prime>\tau_2>\tau_2^\prime>...>\tau_n>\tau_n^\prime>0$
Besides, the integration over time points in the expression for the
n-th perturbative term Eq.~(\ref{gexprG}) can be replaced as follows
\begin{eqnarray}
\label{intrep}
&&\int_0^td\tau_1\int_0^td\tau_1^\prime\int_0^td\tau_2\int_0^td\tau_2^\prime...\int_0^td\tau_n\int_0^td\tau_n^\prime
\rightarrow \nonumber
\\
&&\int_0^td\tau_1\int_0^{\tau_1}d\tau_1^\prime\int_0^{\tau_1^\prime}d\tau_2\int_0^{\tau_2}d\tau_2^\prime...\int_0^{\tau_{n-1}^\prime}d\tau_n\int_0^{\tau_n}d\tau_n^\prime~.
\end{eqnarray}

We can connect every pair of vertices in the diagram of n-th order
by the wiggly $\Gamma$-lines by $(2n)!$ different ways because the
number of vertices is $2n$. But since only the diagrams with
positive direction of time in each propagator $\Gamma(s-s\prime)~,
s>s^\prime$ will bring the non-zero contribution, the whole number
of all diagrams $L_n$ of n-th order is $L_n=(2n)!/2^n$. On the other
hand side, we have $L_n=C^{2n}_2C^{2n-2}_2...C^{4}_2$, where
$C^m_k=\frac{m!}{k!(m-k)!}$ is the number of all possible choices of
pairs of vertices to be connected by the propagator $\Gamma$. Next,
it is clear that in the sum of n-th order diagrams one can meet
identical graphs which can be obtained from each other by permuting
wiggly lines. Thus one can separate all diagrams giving different
contribution as $L_n=D_nn!$, where $D_n$ denotes the number of all
topologically different graphs, and the factor $n!$ is due to the
permutation of every pairs of vertices connected by wiggly line.

Now let us write down the terms of the second ($D_2=3$) and the
third ($D_3=15$) orders in diagrammatic representation in accordance
with the rules established above

\vspace{0.2cm}
\begin{eqnarray}
&&{\mathcal G}_{\bf p}^{(2)}(t)={\mathcal G}_{\bf p}^{(1)}(t)+
\parbox{55mm}{
\begin{fmffile}{diag224}
\begin{fmfgraph}(150,-0)
\fmfleft{o}  \fmfright{i}\fmf{plain}{i,v1} \fmf{plain}{v1,v2}
\fmf{plain}{v2,v3}  \fmf{plain}{v3,v4} \fmf{plain}{v4,o}
\fmffreeze
\fmf{wiggly,right}{v1,v2}\fmf{wiggly,right}{v3,v4}
\fmfdot{v1,v2,v3,v4}
\end{fmfgraph}
\end{fmffile}
}+~
\parbox{55mm}{
\begin{fmffile}{diag1197}
\begin{fmfgraph}(150,-0)
\fmfleft{o}  \fmfright{i} \fmf{plain}{i,v1} \fmf{plain}{v1,v2}
\fmf{plain}{v2,v3}  \fmf{plain}{v3,v4} \fmf{plain}{v4,o}  \fmffreeze
\fmf{wiggly,right}{v1,v3} \fmf{wiggly,right}{v2,v4}
\fmfdot{v1,v2,v3,v4} \fmfright{i}
\end{fmfgraph}
\end{fmffile}
}\nonumber \\
\nonumber
\\
\nonumber \\
 &&~~~~~~~~~~~~ +~\parbox{55mm}{
\begin{fmffile}{diag317}
\begin{fmfgraph}(150,-0)
\fmfleft{o} \fmfright{i} \fmf{plain}{i,v1} \fmf{plain}{v1,v2}
\fmf{plain}{v2,v3} \fmf{plain}{v3,v4} \fmf{plain}{v4,o} \fmffreeze
 \fmf{wiggly,right}{v1,v4}\fmf{wiggly,right}{v2,v3}
 \fmfdot{v1,v2,v3,v4}
\end{fmfgraph}
\end{fmffile}
}
\end{eqnarray}

\begin{eqnarray}
&&{\mathcal G}_{\bf p}^{(3)}(t)={\mathcal G}_{\bf p}^{(2)}(t)-~
\parbox{55mm}{
\begin{fmffile}{diag30033}
\begin{fmfgraph}(200,-20)
\fmfleft{o} \fmfright{i} \fmf{plain}{i,v1} \fmf{plain}{v1,v2}
\fmf{plain}{v2,v3}  \fmf{plain}{v3,v4} \fmf{plain}{v4,v5}
 \fmf{plain}{v5,v6}\fmf{plain}{v6,o}
\fmffreeze \fmf{wiggly,right}{v1,v2} \fmf{wiggly,right}{v3,v4}
\fmfdot{v1,v2,v3,v4,v5,v6}\fmf{wiggly,right}{v5,v6}
\end{fmfgraph}
\end{fmffile}
}
\\
\nonumber
\\
\nonumber
\\
 &&-~
\parbox{72mm}{
\begin{fmffile}{diag300421}
\begin{fmfgraph}(200,-20)
\fmfleft{o} \fmfright{i} \fmf{plain}{i,v1} \fmf{plain}{v1,v2}
 \fmf{plain}{v2,v3}
 \fmf{plain}{v3,v4}
\fmf{plain}{v4,v5}  \fmf{plain}{v5,v6} \fmf{plain}{v6,o} \fmffreeze
\fmf{wiggly,right}{v1,v3}
\fmf{wiggly,right}{v2,v4}\fmf{wiggly,right}{v5,v6}
\fmfdot{v1,v2,v3,v4,v5,v6}
\end{fmfgraph}
\end{fmffile}
} -~\parbox{72mm}{
\begin{fmffile}{diag300431}
\begin{fmfgraph}(200,-20)
\fmfleft{i} \fmfright{o} \fmf{plain}{i,v1} \fmf{plain}{v1,v2}
\fmf{plain}{v2,v3}  \fmf{plain}{v3,v4} \fmf{plain}{v4,v5}
 \fmf{plain}{v5,v6}
 \fmf{plain}{v6,o}
\fmffreeze
\fmf{wiggly,left}{v2,v4}\fmf{wiggly,left}{v5,v6}\fmf{wiggly,left}{v1,v3}
 \fmfdot{v1,v2,v3,v4,v5,v6}
\end{fmfgraph}
\end{fmffile}
} \nonumber
\\
\nonumber
\\
\nonumber
\\
 &&-~
\parbox{72mm}{
\begin{fmffile}{diag30052}
\begin{fmfgraph}(200,-20)
\fmfleft{o} \fmfright{i}\fmf{plain}{i,v1} \fmf{plain}{v1,v2}
\fmf{plain}{v2,v3}  \fmf{plain}{v3,v4} \fmf{plain}{v4,v5}
\fmf{plain}{v5,v6} \fmf{plain}{v6,o} \fmffreeze
 \fmf{wiggly,right}{v1,v4}\fmf{wiggly,right}{v2,v3}
\fmf{wiggly,right}{v5,v6}\fmfdot{v1,v2,v3,v4,v5,v6}
\end{fmfgraph}
\end{fmffile}
}-~
\parbox{72mm}{
\begin{fmffile}{diag30062}
\begin{fmfgraph}(200,-20)
\fmfleft{i}\fmfright{o} \fmf{plain}{i,v1} \fmf{plain}{v1,v2}
\fmf{plain}{v2,v3}  \fmf{plain}{v3,v4} \fmf{plain}{v4,v5}
\fmf{plain}{v5,v6} \fmf{plain}{v6,o} \fmffreeze
\fmf{wiggly,left}{v1,v4}\fmf{wiggly,left}{v2,v3}
\fmf{wiggly,left}{v5,v6}\fmfdot{v1,v2,v3,v4,v5,v6}
\end{fmfgraph}
\end{fmffile}
} \nonumber
\\
\nonumber
\\
\nonumber
\\
\nonumber
\\
 &&-~
\parbox{72mm}{
\begin{fmffile}{diag300791}
\begin{fmfgraph}(200,-20)
\fmfleft{o} \fmfright{i} \fmf{plain}{i,v1} \fmf{plain}{v1,v2}
\fmf{plain}{v2,v3} \fmf{plain}{v3,v4} \fmf{plain}{v4,v5}
\fmf{plain}{v5,v6} \fmf{plain}{v6,o} \fmffreeze
 \fmf{wiggly,right}{v1,v6}
\fmf{wiggly,right}{v2,v5} \fmf{wiggly,right}{v3,v4}
\fmfdot{v1,v2,v3,v4,v5,v6}
\end{fmfgraph}
\end{fmffile}
}-~
\parbox{72mm}{
\begin{fmffile}{diag30092}
\begin{fmfgraph}(200,-20)
\fmfleft{o} \fmfright{i} \fmf{plain}{i,v1} \fmf{plain}{v1,v2}
\fmf{plain}{v2,v3}  \fmf{plain}{v3,v4}
 \fmf{plain}{v4,v5} \fmf{plain}{v5,v6}
\fmf{plain}{v6,o} \fmffreeze
\fmf{wiggly,right}{v2,v3}\fmf{wiggly,right}{v1,v6}
\fmf{wiggly,right}{v4,v5}
 \fmfdot{v1,v2,v3,v4,v5,v6}
\end{fmfgraph}
\end{fmffile}
}\nonumber
\\
\nonumber
\\
\nonumber
\\
\nonumber
\\
 &&-~
\parbox{72mm}{
\begin{fmffile}{diag30083}
\begin{fmfgraph}(200,-20)
\fmfleft{o}  \fmfright{i}\fmf{plain}{i,v1} \fmf{plain}{v1,v2}
\fmf{plain}{v2,v3} \fmf{plain}{v3,v4}
 \fmf{plain}{v4,v5} \fmf{plain}{v5,v6}
\fmf{plain}{v6,o} \fmffreeze \fmf{wiggly,right}{v1,v6}
\fmf{wiggly,right}{v2,v4} \fmf{wiggly,right}{v3,v5}
\fmfdot{v1,v2,v3,v4,v5,v6}
\end{fmfgraph}
\end{fmffile}
}-~
\parbox{72mm}{
\begin{fmffile}{diag300122}
\begin{fmfgraph}(200,-20)
\fmfleft{o}  \fmfright{i}\fmf{plain}{i,v1} \fmf{plain}{v1,v2}
\fmf{plain}{v2,v3}  \fmf{plain}{v3,v4}  \fmf{plain}{v4,v5}
\fmf{plain}{v5,v6} \fmf{plain}{v6,o} \fmffreeze
\fmf{wiggly,right}{v4,v6} \fmf{wiggly,right}{v2,v5}
\fmf{wiggly,right}{v1,v3} \fmfdot{v1,v2,v3,v4,v5,v6}
\end{fmfgraph}
\end{fmffile}
} \nonumber
\\
\nonumber
\\
\nonumber
\\
 &&-~
\parbox{72mm}{
\begin{fmffile}{diag3001395}
\begin{fmfgraph}(200,-20)
\fmfleft{o} \fmfright{i}\fmf{plain}{i,v1} \fmf{plain}{v1,v2}
\fmf{plain}{v2,v3} \fmf{plain}{v4,v3} \fmf{plain}{v4,v5}
\fmf{plain}{v5,v6} \fmf{plain}{v6,o} \fmffreeze
\fmf{wiggly,right}{v1,v4} \fmf{wiggly,right}{v3,v5}
\fmf{wiggly,right}{v2,v6} \fmfdot{v1,v2,v3,v4,v5,v6}
\end{fmfgraph}
\end{fmffile}
} -~
\parbox{72mm}{
\begin{fmffile}{diag30014991}
\begin{fmfgraph}(200,-20)
\fmfleft{o} \fmfright{i}\fmf{plain}{i,v1} \fmf{plain}{v1,v2}
\fmf{plain}{v2,v3}\fmf{plain}{v3,v4} \fmf{plain}{v4,v5}
\fmf{plain}{v5,v6} \fmf{plain}{v6,o} \fmffreeze
 \fmfdot{v1,v2,v3,v4,v5,v6}
 \fmf{wiggly,right}{v2,v4}
\fmf{wiggly,right}{v3,v6} \fmf{wiggly,right}{v1,v5}
\end{fmfgraph}
\end{fmffile}
} \nonumber
\\
\nonumber
\\
\nonumber
\\
 &&-~
\parbox{72mm}{
\begin{fmffile}{diag300151}
\begin{fmfgraph}(200,-20)
\fmfleft{o} \fmfright{i}\fmf{plain}{i,v1} \fmf{plain}{v1,v2}
\fmf{plain}{v2,v3}\fmf{plain}{v3,v4} \fmf{plain}{v4,v5}
\fmf{plain}{v5,v6} \fmf{plain}{v6,o} \fmffreeze
 \fmfdot{v1,v2,v3,v4,v5,v6}
 \fmf{wiggly,right}{v1,v3}
\fmf{wiggly,right}{v2,v6} \fmf{wiggly,right}{v4,v5}
\end{fmfgraph}
\end{fmffile}
}-~
\parbox{72mm}{
\begin{fmffile}{diag300152}
\begin{fmfgraph}(200,-20)
\fmfleft{o} \fmfright{i}\fmf{plain}{i,v1} \fmf{plain}{v1,v2}
\fmf{plain}{v2,v3}\fmf{plain}{v3,v4} \fmf{plain}{v4,v5}
\fmf{plain}{v5,v6} \fmf{plain}{v6,o} \fmffreeze
 \fmfdot{v1,v2,v3,v4,v5,v6}
 \fmf{wiggly,right}{v1,v5}
\fmf{wiggly,right}{v2,v3} \fmf{wiggly,right}{v4,v6}
\end{fmfgraph}
\end{fmffile}
} \nonumber
\\
\nonumber
\\
\nonumber
\\
 &&-~
\parbox{72mm}{
\begin{fmffile}{diag300161}
\begin{fmfgraph}(200,-20)
\fmfleft{o} \fmfright{i}\fmf{plain}{i,v1} \fmf{plain}{v1,v2}
\fmf{plain}{v2,v3}\fmf{plain}{v3,v4} \fmf{plain}{v4,v5}
\fmf{plain}{v5,v6} \fmf{plain}{v6,o} \fmffreeze
 \fmfdot{v1,v2,v3,v4,v5,v6}
 \fmf{wiggly,right}{v1,v5}
\fmf{wiggly,right}{v2,v6} \fmf{wiggly,right}{v3,v4}
\end{fmfgraph}
\end{fmffile}
}-~
\parbox{72mm}{
\begin{fmffile}{diag300162}
\begin{fmfgraph}(200,-20)
\fmfleft{o} \fmfright{i}\fmf{plain}{i,v1} \fmf{plain}{v1,v2}
\fmf{plain}{v2,v3}\fmf{plain}{v3,v4} \fmf{plain}{v4,v5}
\fmf{plain}{v5,v6} \fmf{plain}{v6,o} \fmffreeze
 \fmfdot{v1,v2,v3,v4,v5,v6}
 \fmf{wiggly,right}{v1,v4}
\fmf{wiggly,right}{v2,v5} \fmf{wiggly,right}{v3,v6}
\end{fmfgraph}
\end{fmffile}
}\nonumber
\end{eqnarray}

The last step is to separate all irreducible diagrams, i.e. diagrams
which can not be split into two diagrams of lower order by cutting
one plain inner line, in all orders of perturbation. Let us define
the self-energy graph as the sum of all irreducible blocks of each
perturbation order, i.e.
\begin{eqnarray}
\label{seoexp} &&{\it\Sigma}_{\bf p}(\tau-\tau^\prime)=
\parbox{37mm}{
\begin{fmffile}{diag9082}
\begin{fmfgraph}(100,-0)
\fmfleft{o} \fmf{dashes}{i,v1} \fmf{dashes}{v1,o}
\fmfblob{0.2w}{v1}\fmfright{i}
\end{fmfgraph}
\end{fmffile}
}=-~\parbox{20mm}{
\begin{fmffile}{diag048}
\begin{fmfgraph}(50,-0)
\fmfleft{v2}  \fmf{plain}{v1,v2} \fmf{wiggly,right}{v1,v2}
\fmfdot{v1,v2} \fmfright{v1}
\end{fmfgraph}
\end{fmffile}
}+~\parbox{27mm}{
\begin{fmffile}{diag1199}
\begin{fmfgraph}(68,-0)
\fmfleft{v4}  \fmfright{v1} \fmf{plain}{v1,v2} \fmf{plain}{v2,v3}
\fmf{plain}{v3,v4}  \fmffreeze \fmf{wiggly,right}{v1,v3}
\fmf{wiggly,right}{v2,v4} \fmfdot{v1,v2,v3,v4} \fmfright{i}
\end{fmfgraph}
\end{fmffile}
}+~\parbox{27mm}{
\begin{fmffile}{diag318}
\begin{fmfgraph}(68,-0)
\fmfleft{v4} \fmfright{v1}  \fmf{plain}{v1,v2} \fmf{plain}{v2,v3}
\fmf{plain}{v3,v4}  \fmffreeze
 \fmf{wiggly,right}{v1,v4}\fmf{wiggly,right}{v2,v3}
 \fmfdot{v1,v2,v3,v4}
\end{fmfgraph}
\end{fmffile}
} \nonumber
\\
\nonumber
\\
\nonumber
\\
 &&-~\parbox{40mm}{
\begin{fmffile}{diag3007913}
\begin{fmfgraph}(100,-0)
\fmfleft{v6} \fmfright{v1} \fmf{plain}{v1,v2} \fmf{plain}{v2,v3}
\fmf{plain}{v3,v4} \fmf{plain}{v4,v5} \fmf{plain}{v5,v6} \fmffreeze
 \fmf{wiggly,right}{v1,v6}
\fmf{wiggly,right}{v2,v5} \fmf{wiggly,right}{v3,v4}
\fmfdot{v1,v2,v3,v4,v5,v6}
\end{fmfgraph}
\end{fmffile}
} -~~
\parbox{40mm}{
\begin{fmffile}{diag300922}
\begin{fmfgraph}(100,-0)
\fmfleft{v6} \fmfright{v1} \fmf{plain}{v1,v2} \fmf{plain}{v2,v3}
\fmf{plain}{v3,v4}
 \fmf{plain}{v4,v5} \fmf{plain}{v5,v6}
 \fmffreeze
\fmf{wiggly,right}{v2,v3}\fmf{wiggly,right}{v1,v6}
\fmf{wiggly,right}{v4,v5}
 \fmfdot{v1,v2,v3,v4,v5,v6}
\end{fmfgraph}
\end{fmffile}
}-~~
\parbox{40mm}{
\begin{fmffile}{diag300832}
\begin{fmfgraph}(100,-0)
\fmfleft{v6}  \fmfright{v1} \fmf{plain}{v1,v2} \fmf{plain}{v2,v3}
\fmf{plain}{v3,v4}
 \fmf{plain}{v4,v5} \fmf{plain}{v5,v6}
 \fmffreeze \fmf{wiggly,right}{v1,v6}
\fmf{wiggly,right}{v2,v4} \fmf{wiggly,right}{v3,v5}
\fmfdot{v1,v2,v3,v4,v5,v6}
\end{fmfgraph}
\end{fmffile}
} \nonumber
\\
\nonumber
 \\
 \nonumber
 \\
&& -~~
\parbox{40mm}{
\begin{fmffile}{diag3001222}
\begin{fmfgraph}(100,-0)
\fmfleft{v6}  \fmfright{v1} \fmf{plain}{v1,v2} \fmf{plain}{v2,v3}
\fmf{plain}{v3,v4}  \fmf{plain}{v4,v5} \fmf{plain}{v5,v6}
 \fmffreeze \fmf{wiggly,right}{v4,v6}
\fmf{wiggly,right}{v2,v5} \fmf{wiggly,right}{v1,v3}
\fmfdot{v1,v2,v3,v4,v5,v6}
\end{fmfgraph}
\end{fmffile}
}-~~
\parbox{40mm}{
\begin{fmffile}{diag3001396}
\begin{fmfgraph}(100,-0)
\fmfleft{v6} \fmfright{v1} \fmf{plain}{v1,v2} \fmf{plain}{v2,v3}
\fmf{plain}{v4,v3} \fmf{plain}{v4,v5} \fmf{plain}{v5,v6} \fmffreeze
\fmf{wiggly,right}{v1,v4} \fmf{wiggly,right}{v3,v5}
\fmf{wiggly,right}{v2,v6} \fmfdot{v1,v2,v3,v4,v5,v6}
\end{fmfgraph}
\end{fmffile}
} -~~
\parbox{40mm}{
\begin{fmffile}{diag30014992}
\begin{fmfgraph}(100,-0)
\fmfleft{v6} \fmfright{v1} \fmf{plain}{v1,v2}
\fmf{plain}{v2,v3}\fmf{plain}{v3,v4} \fmf{plain}{v4,v5}
\fmf{plain}{v5,v6}  \fmffreeze
 \fmfdot{v1,v2,v3,v4,v5,v6}
 \fmf{wiggly,right}{v2,v4}
\fmf{wiggly,right}{v3,v6} \fmf{wiggly,right}{v1,v5}
\end{fmfgraph}
\end{fmffile}
} \nonumber
\\
\nonumber
\\
\nonumber
\\
 &&-~~
\parbox{40mm}{
\begin{fmffile}{diag300155}
\begin{fmfgraph}(100,-0)
\fmfleft{v6} \fmfright{v1} \fmf{plain}{v1,v2}
\fmf{plain}{v2,v3}\fmf{plain}{v3,v4} \fmf{plain}{v4,v5}
\fmf{plain}{v5,v6} \fmffreeze
 \fmfdot{v1,v2,v3,v4,v5,v6}
 \fmf{wiggly,right}{v1,v3}
\fmf{wiggly,right}{v2,v6} \fmf{wiggly,right}{v4,v5}
\end{fmfgraph}
\end{fmffile}
}-~~
\parbox{40mm}{
\begin{fmffile}{diag3001531}
\begin{fmfgraph}(100,-0)
\fmfleft{v6} \fmfright{v1} \fmf{plain}{v1,v2}
\fmf{plain}{v2,v3}\fmf{plain}{v3,v4} \fmf{plain}{v4,v5}
\fmf{plain}{v5,v6}  \fmffreeze
 \fmfdot{v1,v2,v3,v4,v5,v6}
 \fmf{wiggly,right}{v1,v5}
\fmf{wiggly,right}{v2,v3} \fmf{wiggly,right}{v4,v6}
\end{fmfgraph}
\end{fmffile}
} -~~
\parbox{40mm}{
\begin{fmffile}{diag300165}
\begin{fmfgraph}(100,-0)
\fmfleft{v6} \fmfright{v1} \fmf{plain}{v1,v2}
\fmf{plain}{v2,v3}\fmf{plain}{v3,v4} \fmf{plain}{v4,v5}
\fmf{plain}{v5,v6}  \fmffreeze
 \fmfdot{v1,v2,v3,v4,v5,v6}
 \fmf{wiggly,right}{v1,v5}
\fmf{wiggly,right}{v2,v6} \fmf{wiggly,right}{v3,v4}
\end{fmfgraph}
\end{fmffile}
} \nonumber \\
\nonumber \\
\nonumber
\\
 &&-~~
\parbox{40mm}{
\begin{fmffile}{diag300164}
\begin{fmfgraph}(100,-0)
\fmfleft{v6} \fmfright{v1} \fmf{plain}{v1,v2}
\fmf{plain}{v2,v3}\fmf{plain}{v3,v4} \fmf{plain}{v4,v5}
\fmf{plain}{v5,v6} \fmffreeze
 \fmfdot{v1,v2,v3,v4,v5,v6}
 \fmf{wiggly,right}{v1,v4}
\fmf{wiggly,right}{v2,v5} \fmf{wiggly,right}{v3,v6}
\end{fmfgraph}
\end{fmffile}
}+~...~.
\end{eqnarray}

Here the time points $\tau$ and $\tau^\prime$ in
$\Sigma(\tau-\tau^\prime)$ denote the left and the right vertices of
each irreducible block on the right hand side of equation,
respectively. Then on can rewrite the expansion for the Green's
function ${\mathcal G}_{\bf p}(t)$ with the help of the self-energy
operator as follows
\begin{eqnarray}
\label{gfse}
 {\mathcal G}_{\bf p}(t)&=&
 \parbox{15mm}{
\begin{fmffile}{diag725011}
\begin{fmfgraph}(40,-0)
\fmfleft{o} \fmf{plain}{i,o} \fmfright{i}
\end{fmfgraph}
\end{fmffile}}+~
\parbox{37mm}{
\begin{fmffile}{diag900235}
\begin{fmfgraph}(100,-0)
\fmfleft{o} \fmf{plain}{i,v1} \fmf{plain}{v1,o}
\fmfblob{0.2w}{v1}\fmfright{i}\fmf{phantom}{i,o}
\end{fmfgraph}
\end{fmffile}
}+~
\parbox{37mm}{
\begin{fmffile}{diag90221}
\begin{fmfgraph}(100,-0)
\fmfleft{o} \fmf{plain}{i,v1}\fmf{plain}{v1,v2} \fmf{plain}{v2,o}
\fmfblob{0.2w}{v1}\fmfblob{0.2w}{v2}\fmfright{i}
\end{fmfgraph}
\end{fmffile}
}+~\parbox{37mm}{
\begin{fmffile}{diag902216}
\begin{fmfgraph}(100,-0)
\fmfleft{o} \fmf{plain}{i,v1}\fmf{plain}{v1,v2}
\fmf{plain}{v2,v3}\fmf{plain}{v3,o}
\fmfblob{0.2w}{v1}\fmfblob{0.2w}{v2}\fmfblob{0.2w}{v3}\fmfright{i}
\end{fmfgraph}
\end{fmffile}
}
\nonumber \\
\nonumber \\
  &+&~...~+~
\parbox{37mm}{
\begin{fmffile}{diag90221}
\begin{fmfgraph}(100,-0)
\fmfleft{o} \fmf{plain}{i,v1}\fmf{plain}{v1,v2} \fmf{plain}{v2,o}
\fmfblob{0.2w}{v1}\fmfblob{0.2w}{v2}\fmfright{i}
\end{fmfgraph}
\end{fmffile}
}\cdot\cdot\cdot\parbox{37mm}{
\begin{fmffile}{diag902114}
\begin{fmfgraph}(100,-0)
\fmfleft{o} \fmf{plain}{i,v1} \fmf{plain}{v1,o}\fmf{phantom}{v1,o}
\fmfblob{0.2w}{v1}\fmfright{i}
\end{fmfgraph}
\end{fmffile}
}+~...~\nonumber
\\
\nonumber
\\
&=&\frac{1}{~~\parbox{15mm}{
\begin{fmffile}{diag725}
\begin{fmfgraph}(38,-0)
\fmfleft{o} \fmf{plain,tension=0.5}{i,o} \fmfright{i}
\end{fmfgraph}
\end{fmffile}}^{{\bf -1}}
~~-~~\parbox{15mm}{\begin{fmffile}{diag825}
\begin{fmfgraph}(38,-0)
\fmfleft{o} \fmf{dashes}{i,v1}
\fmfblob{0.25w}{v1}\fmf{dashes}{v1,o}\fmf{phantom,right}{o,i}\fmfright{i}
\end{fmfgraph}
\end{fmffile}}}~.
\end{eqnarray}

Due to the replacement (\ref{intrep}) each term of perturbation is
now represented by a convolution of the free propagators $G_{\bf p}$
and the self-energy operators $\Sigma_{\bf p}$, i.e.
\begin{eqnarray}
&&\parbox{37mm}{
\begin{fmffile}{diag9022102}
\begin{fmfgraph}(100,-0)
\fmfleft{o} \fmf{dashes}{i,v1}\fmf{plain}{v1,v2} \fmf{plain}{v2,o}
\fmfblob{0.2w}{v1}\fmfblob{0.2w}{v2}\fmfright{i}
\end{fmfgraph}
\end{fmffile}
}\cdot\cdot\cdot\parbox{37mm}{
\begin{fmffile}{diag90211403}
\begin{fmfgraph}(100,-0)
\fmfleft{o} \fmf{plain}{i,v1} \fmf{dashes}{v1,o}\fmf{phantom}{v1,o}
\fmfblob{0.2w}{v1}\fmfright{i}
\end{fmfgraph}
\end{fmffile}
}~=\int_0^td\tau_1G_{\bf
p}(t-\tau_1)\int_0^{\tau_1}d\tau_2{\it\Sigma}_{\bf p}(\tau_1-\tau_2)
\nonumber
\\
\nonumber
\\
 &&\times\int_0^{\tau_2}d\tau_3G_{\bf
p}(\tau_2-\tau_3)\int_0^{\tau_3}d\tau_4{\it\Sigma}_{\bf
p}(\tau_3-\tau_4)...\int_0^{\tau_{k-1}}d\tau_k{\it\Sigma}_{\bf
p}(\tau_{k-1}-\tau_{k})G_{\bf p}(\tau_k)~.
\end{eqnarray}
So for the Laplace transform components the equation (\ref{gfse})
reads
\begin{eqnarray}
\label{gfd}
 {\bar{\mathcal G}}_{\bf p}(\omega)&=&{\bar G}_{\bf
p}(\omega)+{\bar G}_{\bf p}(\omega){\bar{\it\Sigma}}_{\bf
p}(\omega){\bar G}_{\bf p}(\omega)+...+{\bar G}_{\bf
p}(\omega){\bar{\it\Sigma}}_{\bf p}(\omega){\bar G}_{\bf
p}(\omega)...{\bar{\it\Sigma}}_{\bf p}(\omega){\bar G}_{\bf
p}(\omega)+...\nonumber
\\
&=&\frac{1}{{\bar G}_{\bf p}^{-1}(\omega)-{\bar{\it\Sigma}}_{\bf
p}(\omega)}~.
\end{eqnarray}
Here the bar denotes Laplace transform. Thus at this point we
obtained the Dyson's-type equation for the Green's function
Eq.~(\ref{gfd}). The poles of the Green's function are determined by
the solution of the spectral equation ${\bar G}_{\bf
p}^{-1}(\omega)-{\bar{\it\Sigma}}_{\bf p}(\omega)=0$. The
contribution of the self-energy operator into the spectral equation
can be interpreted as the influence of the cloud of the virtual BEC
excitations surrounding the particle. The solution of this equation
will be discussed in the next section.

\section{DISPERSION RELATION AND THE ENERGY SPECTRUM OF AN IMPURITY}
In the previous section we obtained the expansion for the
self-energy operator in powers of coupling constant
\begin{eqnarray}
{\it\Sigma}_{\bf p}(\tau)=\sigma^{(1)}_{\bf
p}(\tau)+\sigma^{(2)}_{\bf p}(\tau)+...~,
\end{eqnarray}
where the n-th expansion term $\sigma^{(n)}_{\bf p}(\tau)$ consists
of all diagrams with n wiggly lines in accordance with the expansion
(\ref{seoexp}) and is proportional to $g^{2n}$. In accordance with
Eq.~(\ref{gfd}) the Green's function yields
\begin{eqnarray}
{\bar{\mathcal G}}_{\bf p}(\omega)=\frac{1}{\omega+iE({\bf
p})-{\bar{\it\Sigma}}_{\bf p}(\omega)}~,
\end{eqnarray}
and its inverse Laplace transform is given by the Fourier-Mellin
integral
\begin{eqnarray}
{\mathcal G}_{\bf p}(t)=\lim_{\varepsilon\to 0+}\frac{1}{2\pi
i}\int_{-i\infty+\varepsilon}^{i\infty+\varepsilon}{\bar{\mathcal
G}}_{\bf p}(\omega)e^{\omega t}d\omega~.
\end{eqnarray}
Replacing the integration variable in the above integral as
$\omega=-i\Omega$, for the origin of the Green's function one gets
\begin{eqnarray}
\label{gfgr} {\mathcal G}_{\bf p}(t)=\frac{1}{2\pi i}\int
d\Omega\frac{e^{-i\Omega t}}{E({\bf
p})-\Omega+i{\bar{\it\Sigma}}_{\bf p}(\Omega)}~,
\end{eqnarray}
where the integration is performed over the contour shown on
Fig.~(\ref{contour}), and ${\bar{\it\Sigma}}_{\bf
p}(\Omega)={\bar{\it\Sigma}}_{\bf p}(\omega=-i\Omega)$.
\begin{figure}
\includegraphics[width=10.0cm]{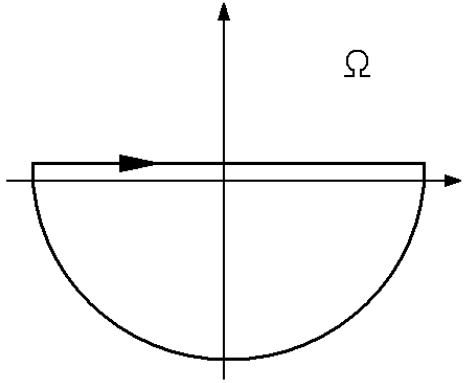}
\caption{ \label{contour} Integration contour for the calculation of
inverse Laplace transform of the Green's function. }
\end{figure}
First, let us concentrate on the lowest order perturbative results
that can be obtained by taking into account the first expansion term
with one loop in the expression for the self energy operator
Eq.~(\ref{seoexp})
\vspace{0.5cm}
\begin{eqnarray}
\label{seo1} \sigma_{\bf p}^{(1)}(\tau-\tau^\prime)=-~\parbox{20mm}{
\begin{fmffile}{diag048}
\begin{fmfgraph}(50,-0)
\fmfleft{v2}  \fmf{plain}{v1,v2} \fmf{wiggly,right}{v1,v2}
\fmfdot{v1,v2} \fmfright{v1}
\end{fmfgraph}
\end{fmffile}
} =-\sum_{\bf k}\gamma_{\bf k}^2G_{\bf
p-k}(\tau-\tau^\prime)\Gamma_{\bf k}(\tau-\tau^\prime)~.
\end{eqnarray}
Evaluating the Laplace transform of Eq.~(\ref{seo1}) and subtituting
it into Eq.~(\ref{gfd}), for the Green's function in one loop
approximation one gets
\begin{eqnarray}
{\mathcal G}^{(1)}_{\bf p}(t)=\frac{1}{2\pi
i}\int_{-\infty}^{\infty}d\Omega\frac{e^{-i\Omega t}}{E({\bf
p})-\Omega+\sum_{\bf k}\frac{\gamma_{\bf k}^2}{\Omega-E({\bf
p-k})-\epsilon({\bf k})+i0}-i0}~.
\end{eqnarray}
The third term in the denominator is significant only if $\Omega$ is
close to $E({\bf p})$. Thus for the pole of the above integral we
have
\begin{eqnarray}
\label{omega1} \Omega_0^{(1)}=E({\bf p})+\sum_{\bf k}{\it
P}\frac{\gamma_{\bf k}^2}{E({\bf p})-E({\bf p-k})-\epsilon({\bf
k})}-i\pi\sum_{\bf k}\delta\big(E({\bf p})-E({\bf
p-k})-\epsilon({\bf k})\big)~.
\end{eqnarray}
As it is well known, the real part of the pole of the Green's
function determines the energy spectrum $E_i({\bf p})={\rm
Re}(\Omega)$ while the imaginary part defines the dissipation rate.
Thus the second term in the right hand side of Eq.~(\ref{omega1})
represents the correction to the energy of the impurity due to the
interaction with BEC while the last term describes the dissipation
process, i.e. the energy transfer between impurity and BEC.

Let us calculate the energy of the particle at zero momentum ${\bf
p}=0$. After performing the thermodynamic limit
($N\to\infty,~~V\to\infty,~~N/V=n$), i.e. replacing the sum over
momenta $\sum_{\bf k}$ by the integral $\frac{V}{(2\pi)^3}\int d{\bf
k}$, the zero point energy can be written as
\begin{eqnarray}
E_i^{(1)}(p=0)=gn-\frac{g^2n}{16\pi^3m}\int d{\bf
k}\frac{k^2}{2m\epsilon(k)}\frac{1}{\epsilon(k)+k^2/2m}~.
\end{eqnarray}
In order to prevent the divergence at large momenta in the above
integral one has to renormalize the coupling constant $g$ by taking
into account the second order Born approximation for the scattering
length $a$
\begin{eqnarray}
g=\frac{2\pi a}{m_r}\left(1+\frac{2a}{\pi}\int dk\right)~,
\end{eqnarray}
where $m_r=(1/m+1/M)^{-1}$ is the reduced mass. The energy
reexpanded in powers of scattering length is now finite and given as
\begin{eqnarray}
E_i^{(1)}(p=0)=\frac{2\pi
an}{m_r}\left(1+\frac{2amc}{\pi}I_0(m/M)\right)~,
\\
I_0(z)=\frac{2z\sqrt{z^2-1}-2\ln\big(z+\sqrt{z^2-1}\big)}{\sqrt{(z^2-1)^3(z^2+1)}}~.
\end{eqnarray}
At this point one can consider the interesting case if the impurity
has the same mass as the condensate particle and the scattering
length $a$ equals the scattering length for the interaction between
Bose particles in the condensate, i.e. $c=\sqrt{4\pi an}/m$. Then
the energy correction reads ($I_0(z=0)=8/3$)
\begin{eqnarray}
E_i^{(1)}(p=0,m=M)=\frac{4\pi
an}{m}\left(1+\frac{32}{3}\sqrt{\frac{a^3n}{\pi}}\right)=\mu_B~.
\end{eqnarray}
So the energy of the resting impurity in this case coincides with
the chemical potential of the interacting Bose gas in Bogoliubov's
approximation (the expression for $\mu_B$ can be obtained using
equation for the ground state energy of BEC (\ref{gsebec}) and can
be found, for example, in Ref.~\cite{landau_v9}).

The probability for the particle to stay in its initial state $|{\bf
p}\rangle$ is given by $w_{\bf p}=|{\mathcal G}_{\bf p}(t)|^2$ and
it decays exponentially like $\exp(-\lambda t)$, where the
transition rate $\lambda$ is given by the imaginary part of the pole
$\Omega_0^{(1)}$, i.e.
\begin{eqnarray}
\label{tranrate}
 \lambda_{\bf p}&=&2\pi\sum_{\bf k}\gamma_{\bf
k}\delta\big(E({\bf p})-E({\bf p-k})-\epsilon({\bf k})\big)
\nonumber \\
 &=&\frac{g^2n}{8\pi^2m}\int d{\bf
k}\frac{k^2}{\epsilon(k)}\delta\big(E({\bf p})-E({\bf
p-k})-\epsilon({\bf k})\big)~.
\end{eqnarray}
The above result corresponds to the Golden rule approximation and
reflects the Landau's criterion for the energy dissipation in BEC
since the integral in the right hand side of Eq.~(\ref{tranrate}) is
not zero only if the momentum of the impurity particle $p$ is more
than its critical value $p_c=Mc$.

Now we turn to the contribution of the next order perturbation term
represented by a couple of two-loop diagrams
\vspace{0.9cm}
\begin{eqnarray}
\label{sigma2}
 {\bar\sigma}_{\bf p}^{(2)}(\Omega)&=& ~~\parbox{27mm}{
\begin{fmffile}{diag1199}
\begin{fmfgraph}(68,-0)
\fmfleft{v4}  \fmfright{v1} \fmf{plain}{v1,v2} \fmf{plain}{v2,v3}
\fmf{plain}{v3,v4}  \fmffreeze \fmf{wiggly,right}{v1,v3}
\fmf{wiggly,right}{v2,v4} \fmfdot{v1,v2,v3,v4} \fmfright{i}
\end{fmfgraph}
\end{fmffile}
}+~~\parbox{27mm}{
\begin{fmffile}{diag318}
\begin{fmfgraph}(68,-0)
\fmfleft{v4} \fmfright{v1}  \fmf{plain}{v1,v2} \fmf{plain}{v2,v3}
\fmf{plain}{v3,v4}  \fmffreeze
 \fmf{wiggly,right}{v1,v4}\fmf{wiggly,right}{v2,v3}
 \fmfdot{v1,v2,v3,v4}
\end{fmfgraph}
\end{fmffile}
}
 \nonumber
\\
\nonumber
\\
&=& \sum_{\bf k,k^\prime}i\gamma_{\bf k}^2\gamma_{\bf
k^\prime}^2\frac{E({\bf p-k})+E({\bf p-k^\prime})+\epsilon({\bf
k})+\epsilon({\bf k^\prime})-2\Omega}{(\epsilon({\bf k})+E({\bf
p-k})-\Omega-i0)^2(\epsilon({\bf k^\prime})+E({\bf
p-k^\prime})-\Omega-i0)} \nonumber
\\
&\times&\frac{1}{(\epsilon({\bf k})+\epsilon({\bf k^\prime})+E({\bf
p-k-k^\prime})-\Omega-i0)}~.
\end{eqnarray}
\begin{figure}
\includegraphics[width=10.0cm]{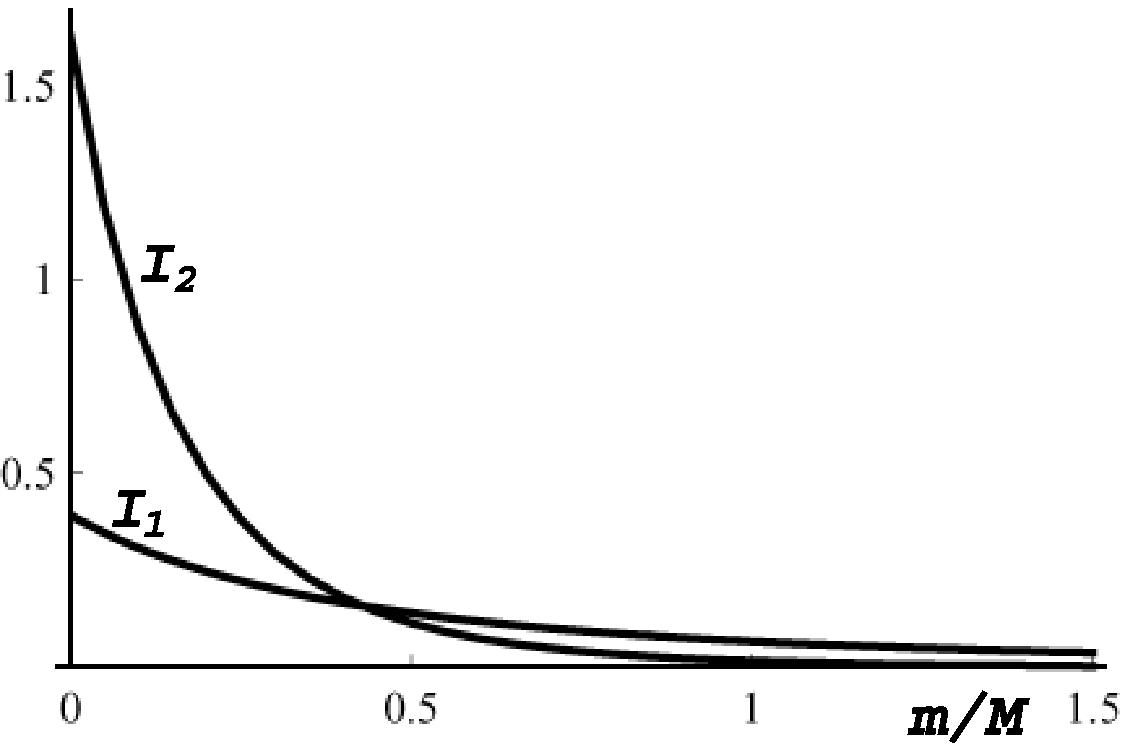}
\caption{ \label{figint} The functions that enter the expression for
the effective mass of an impurity in BEC as given by
Eq.~(\ref{mass})}
\end{figure}
The spectral equation defining the poles of the particle Green's
function is now given by
\begin{eqnarray}
E({\bf p})-\Omega_0+i{\bar \sigma}^{(1)}_{\bf p}(\Omega_0)+i{\bar
\sigma}^{(2)}_{\bf p}(\Omega_0)=0~.
\end{eqnarray}
Since last two terms in the above equation are  small, we will find
the solution $\Omega_0$ iteratively, i.e. in the form of expansion
in powers of coupling constant
\begin{eqnarray}
\label{se2}
 \Omega_0^{(2)}=E({\bf p})+i{\bar \sigma}^{(1)}_{\bf p}(E({\bf
p}))-\frac{\partial{\bar\sigma}^{(1)}_{\bf
p}(\Omega)}{\partial\Omega}\Big|_{E({\bf p})}{\bar
\sigma}^{(1)}_{\bf p}(E({\bf p}))+i{\bar \sigma}^{(2)}_{\bf
p}(E({\bf p}))~.
\end{eqnarray}
The third and the fourth terms in (\ref{se2}) represents the
correction of the order $g^4$ to the result (\ref{omega1}). One can
see that the imaginary part of the pole coming from the denominator
in the right hand side of Eq.~(\ref{sigma2}) and defining the
correction to the transition rate (\ref{tranrate}) does not
contradict with the Landau's criterion. Finally, the pole with the
second order contribution can be written as follows
\begin{eqnarray}
\label{omega2} &&\Omega_0^{(2)}({\bf
p})=\Omega_0^{(1)}+\frac{g^4n^2m^4}{4\pi^6M}\int d{\bf l}d{\bf
l^\prime}\frac{{\bf l}{\bf l^\prime}}{\sqrt{(1+4/l^2)(1+4/l^{\prime
2})}}
\\
\nonumber
\\
&\times&\frac{e(l)+e(l^\prime)-2z{\bf p}({\bf l}+{\bf
l^\prime})/k_c}{\Big(e(l)-2z{\bf p}{\bf
l}/k_c\Big)^2\Big(e(l^\prime)-2z{\bf p}{\bf
l^\prime}/k_c\Big)^2\Big(e(l)+e(l^\prime)+2z{\bf l}{\bf
l^\prime}-2z{\bf p}({\bf l}+{\bf l^\prime})/k_c\Big)}+{\rm
Im}\big(\Omega_0^{(2)}({\bf p})\big)~.\nonumber
\end{eqnarray}
where $e(l)=l^2\big(\sqrt{1+4/l^2}+z\big)$, $k_c=mc$ and $z=m/M$.
The integration in the right hand side of Eq.~(\ref{omega2}) is
performed over dimensionless vectors ${\bf l,~l^\prime}$. The second
order contribution to the zero point energy does not depend on the
speed of sound and diverges logarithmically at large momenta.

Next, we will consider the dissipationless motion of the impurity
with the momentum $p$ less than the critical momentum $Mc$, i.e.
${\rm Im}\big(\Omega_0({\bf p})\big)=0$, and expand the function
$\Omega^{(2)}_0(p)$ up to second order in $p$. In absence of
anisotropy the term linear in $p$ disappears, and the energy of the
impurity can be written in the form
\begin{eqnarray}
E_i^{(2)}({\bf p})=E_i({\bf p}=0)+\frac{p^2}{2M_{ef}^{(2)}}~,
\end{eqnarray}
where the effective mass of the impurity is given by following
expression
\begin{eqnarray}
\label{mass}
M_{ef}^{(2)}=M\left[1-\frac{32}{3}g_MI_1\left(\frac{m}{M}\right)+\frac{8}{3\pi^2}g_M^2I
_2\left(\frac{m}{M}\right)\right]^{-1}~.
\end{eqnarray}
Here we introduced new dimensionless expansion parameter
\begin{eqnarray}
g_M=\frac{a^2n}{p_c}\left(\frac{m}{m_r}\right)^2,~~p_c=Mc~.
\end{eqnarray}
Two dimensionless functions $I_1$ and $I_2$ represent a large
algebraic expressions that contain expansion of the integral of Eq.
~(\ref{omega2}). The evaluation of these functions (numerical for
$I_2$) is shown on Fig.~(\ref{figint}).

\section{CONCLUSION}

In this work we developed a systematic perturbation theory for the
quantum propagator of an impurity in the degenerate BEC.  The
expansion of a Green's function is resummed as an expansion of its
poles by introducing the self energy operator with the help of the
diagrammatic technique. We demonstrate the use of this theory by
computing the first two orders of the correction to the free
propagator. In this way we obtain the energy spectrum and effective
mass of an impurity in BEC. This theory gives access to the
calculation of the properties of realistic systems in which the
impurity-BEC interaction is not necessarily weak.  Given the
ordinary expansion developed in this work one can obtain strong
coupling expansions by employing the variational resummation of the
ordinary perturbation series \cite{kleinertbook}. We believe that
the application of the variational perturbation theory would allow
the nonperturbative calculation of the effective mass as well as
finding the regime of self-localization of an impurity.

\end{document}